\documentclass[fleqn,twoside]{article}
\usepackage{espcrc2}
 
\usepackage{graphicx}
\usepackage[figuresright]{rotating}
 
\newcommand{\AmS}{{\protect\the\textfont2
  A\kern-.1667em\lower.5ex\hbox{M}\kern-.125emS}}
 
\title{\bf Pion Decay Constant, $Z_A$ and Chiral Log from Overlap Fermions}

\author{S.J. Dong\address{Dept. of Physics and Astronomy, University of 
Kentucky, Lexington, KY 40506}, T. Draper$^a$,
I. Horv\'{a}th$^a$,
F.X. Lee\address{Center for Nuclear Studies, Dept. of Physics, George 
Washington University, Washington, DC 20052}\address{Jefferson Lab,
 12000 Jefferson Avenue, Newport News, VA 23606}, 
and J.B. Zhang
\address{ CSSM and Dept. of Physics and Math. Physics, University of Adelaide,
SA 5005, Australia} }

\begin{document}

\begin{abstract}
We report our calculation of the pion decay constant $f_\pi$, the axial
renormalization constant $Z_A$, and the quenched chiral logarithms
from the overlap fermions. The calculation is done on a quenched $20^4$ lattice
at $a=0.148$ fm using tree level tadpole 
improved gauge action. The
smallest pion mass we reach is about 280 MeV. The lattice size is about 4
times the Compton wavelength of the lowest mass pion.
\end{abstract}

\maketitle
\section{Numerical Details}
For Neuberger's overlap fermion\cite{Neuberger}, we adopt the following
form for the massive fermion action
\begin{eqnarray}
D(m_0)&=&(\rho+\frac{m_0a}{2})+(\rho-\frac{m_0a}{2})
\gamma_5\epsilon (H) \\
\epsilon (H) &=& H /\sqrt{H^2} \\
H &=& \gamma_5 D_w
\end{eqnarray}
$D_w$ is the usual Wilson fermion operator, and
\begin{eqnarray}
\rho &=& - (1/2\kappa -4 ) = 1.368 ~~{\rm for~\kappa=0.19}
\end{eqnarray}
We adopt the optimal rational approximation~\cite{Cal} to
approximate the matrix sign function. A deflation algorithm is employed to 
accelerate the nested loop~\cite{Cal}. 
We work on a $20^4$
lattice with $\beta=7.60$ tree level tadpole improved L\"{u}scher-Weisz gauge 
action projecting out 85 smallest eigenmodes of $H$. 
We employed shifted matrix inversion method to
calculate 16 quark masses ranging from $m_0a = 0.01505$
to $m_0a = 0.2736$. As we shall
see, the scale determined from $f_{\pi}$ is $a=0.148$ fm which
makes the physical length of the lattice to be 2.96 fm. The smallest pion mass
turns out to be $\sim 280\,{\rm MeV}$ so that the size of the lattice is about
4.1 times of the Compton wavelength of the lowest mass pion.  

We adopt the periodic boundary condition for the spatial dimensions and the
fixed boundary condition in the time direction so that we can have effectively
a longer range of time separation between the source and sink to examine the
meson propagators with small quark masses.
\section{Zero Mode Effects in Meson Propagators}
The quark zero mode is known to contribute to 
the zero-momentum pseudoscalar propagator such that
\begin{eqnarray}
&{\int}& d^3x \langle \pi(x) \pi(0)\rangle = \nonumber\\
& \int & d^3x [\sum_{i,j = zm}
\frac{tr(\psi^{\dagger}_j(x)\psi_i(x))tr(\psi^{\dagger}_i(0)\psi_j(0))}{m_0^2}
\nonumber\\
&+& 2\sum_{i = 0, \lambda > 0}\frac{tr(\psi^{\dagger}_{\lambda}(x)\psi_i(x))
tr(\psi^{\dagger}_i(0)\psi_{\lambda}(0))}{m_0 (\lambda^2 + m_0^2)}] \nonumber\\
&+&\frac{|\langle 0|\pi(0)|\pi\rangle|^2 e^{- m_{\pi}t}}{2 m_{\pi}}
\end{eqnarray}
The first term is purely the zero-mode contribution. The second term is
the cross term between the zero modes and the non-zero modes. Both zero-mode 
terms 
will go down with volume like $1/\sqrt{V}$ and
are finite volume artifacts.

The pseudoscalar propagator for a light quark mass ($m_0a = 0.01915$) 
in Fig.~1 exhibits a kink at $ t/a \sim 8 - 9$.

\begin{figure}[h]
\includegraphics{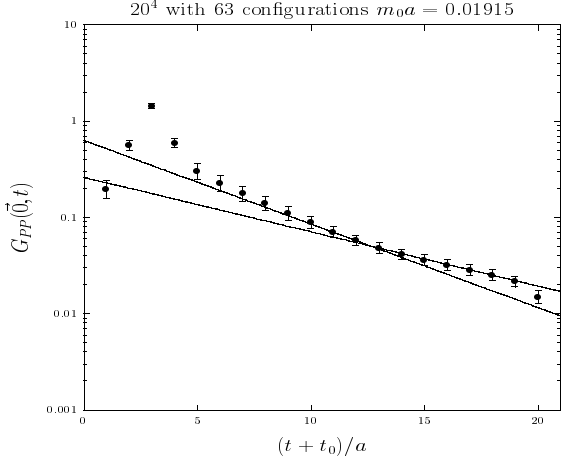}
\vspace{4.6cm}
\caption{Pion propagator for $m_0a = 0.01915$. }
\end{figure}
Zero mode makes the pion propagator fall off faster in $t/a$ at short $t$. 
For this reason, we
turn to the propagator $G_{A_4P}(\vec{p} = 0, t)$ instead for pion mass. 
Since the
zero modes on one gauge configuration have the same chirality, the
pure zero mode contribution vanishes.

\begin{figure}[h]
\includegraphics{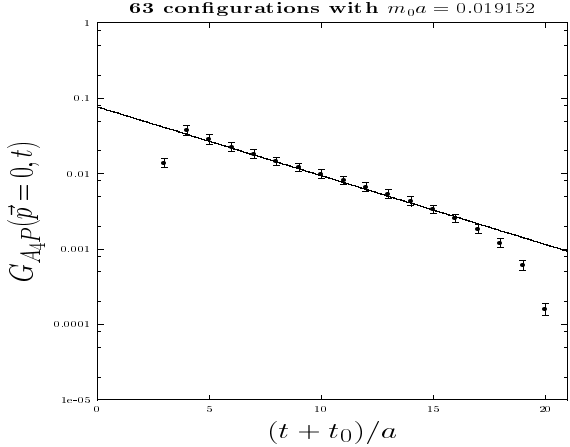}
\vspace{4.4cm}
\caption{Propagator $G_{A_4P}(\vec{p} = 0, t)$ for $m_0 a = 0.01915$ for all
63 configurations.}
\end{figure}

\section{Pion Decay Constant $f_\pi$ and $Z_A$}
We calculate $Z_A$ by the formulas
\begin{eqnarray} 
Z_A &=& \lim_{t \longrightarrow \infty} \frac{ 2 m_0 G_{PP}(\vec{p}= 0, t)}
{G_{\partial_4A_4P}(\vec{p}= 0, t)} \label{ZA1}\\
Z_A &=& \lim_{t \longrightarrow \infty} \frac{ 2 m_0 G_{PP}(\vec{p}= 0, t)}
{m_{\pi} G_{A_4P}(\vec{p}= 0, t)}. \label{ZA}
\end{eqnarray}
The results are shown in the Fig.~3

\begin{figure}[h]
\includegraphics{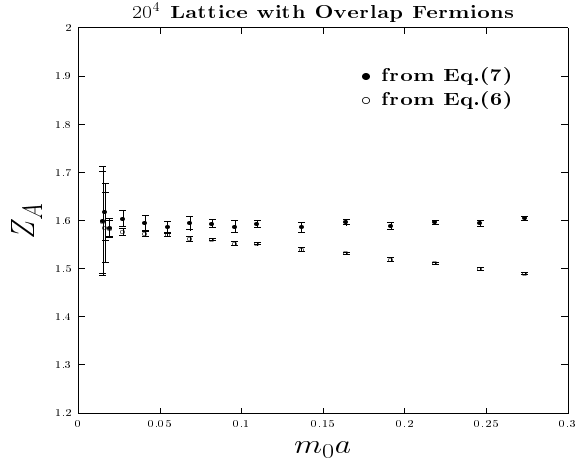}
\vspace{4.5cm}
\caption{$Z_A$ vs quark mass $m_0a$ with two forms}
\end{figure}
The difference is because $G_{\partial_4A_4P}(\vec{p}= 0, t)$ invokes a 
numerical $O(a^2)$ error. Our final value in chiral limit is
\( Z_A=1.589\pm 0.004 \).

 We calculate the pion decay constant by using
\begin{eqnarray}
\lefteqn{f_{\pi}a= \lim_{t \longrightarrow \infty}}
\label{fpi2} \\
&&\frac{2 m_0a\,\sqrt{G_{PP}(\vec{p}=0,t)}
G_{A_4P}(\vec{p}=0,t)}{\sqrt{m_{\pi}a}\,G_{\partial_4A_4P}(\vec{p}=0,t)} 
\cdot e^{m_{\pi}t/2} 
\nonumber
\end{eqnarray}
\begin{equation}
f_{\pi}a=\lim_{t \longrightarrow \infty}\frac{2 m_0a\, 
\sqrt{G_{PP}(\vec{p}= 0, t)
\,m_{\pi}a} e^{m_{\pi}t/2}}{(m_{\pi}a)^2} \label{fpi1}
\end{equation}
The results are shown in Fig.~4. The difference reflects the 
numerical $O(a^2)$ in $G_{\partial_4A_4P}(\vec{p}= 0, t)$. With all 16 data 
points in a linear fit, we find
\begin{equation}
f_{\pi}a = 0.0691(11) + 0.109(56)\,m_0a
\end{equation}
Comparing with the
experimental value $f_{\pi} = 92.4\, {\rm MeV}$, we determine the scale of our
lattice to be $a = 0.148(2)\, {\rm fm}$.

\begin{figure}[th]
\includegraphics{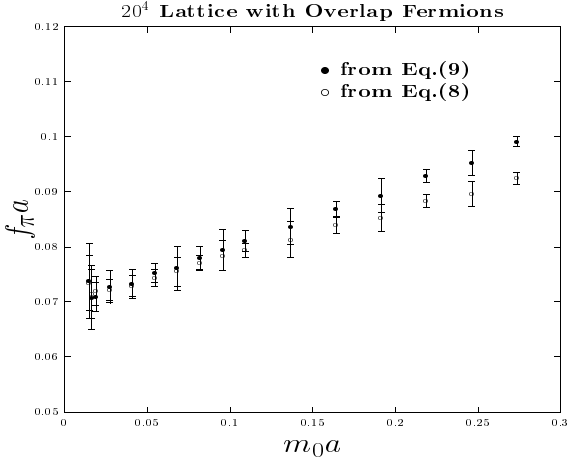}
\vspace{4.5cm}
\caption{Renormalized $f_{\pi}a$ vs quark mass $m_0a$. }
\end{figure}
\section{Quenched Chiral Logs from $f_P$}
In the quenched approximation of QCD, one ignores the virtual quark
loops. The study of the anomalous chiral behavior predicted the chiral-log 
pathologies in
the pseudoscalar meson masses\cite{Log}.  
We look for the chiral-log
in $f_P$ the pseudoscalar decay constant.
\begin{eqnarray}
f_P &=& \langle 0|\bar{\psi} i \gamma_5 \psi |\pi(\vec{p} = 0)\rangle 
\nonumber \\
&=&\lim_{t/a \gg 1} Z_P \sqrt{G_{PP}(t) 2 m_P} e^{ m_Pt/2}
\end{eqnarray}
According to the quenched chiral perturbation theory it should behave as 
$(\frac{1}{m_{\pi}^2})^{\delta}$. For stability we fit the
$f_P$ data in the logarithm  form.
\begin{eqnarray}
\lefteqn{f_P a^2= } \label{f_P_log} \\
&& \tilde{f_P} a^2 \{1 -\delta [\ln(Am_0 a/\Lambda_{\chi}^2 a^2) + 1]\} 
+ B m_0 a  \nonumber
\end{eqnarray}
The results are shown in Fig.~5. Excluding the lowest two mass points from
fitting results in reasonably small errors and $\chi^2/DF$ 
less than unity are in Table 1. 
\begin{table}[h]
\begin{center}
\caption{$\Lambda_{\chi}$ (GeV), 
$\tilde{f_P} a^2$, $\delta$ and $\chi^2/DF$ as fitted
for $f_P a^2$.}
\begin{tabular}{lllll}
\hline
$\Lambda_{\chi}$ & $\tilde{f_P} a^2$  & $\delta$ & B & $\chi^2/DF$  \\
\hline
  0.6   &  0.083(2)  & 0.35(4) & 0.36(2) & 0.93   \\
  0.8   &  0.060(3)  & 0.48(7) & 0.36(2) & 0.93     \\
\hline
\end{tabular}
\end{center}
\end{table}
\begin{figure}[tb]
\includegraphics{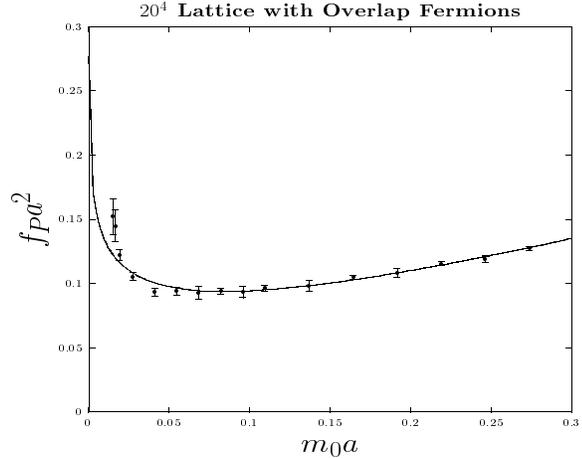}
\vspace{4.5cm}
\caption{Renormalized $f_Pa^2$ vs quark mass $m_0a$. The solid line is a fit
excluding the two smallest quark masses with $\Lambda_{\chi}$ = 0.8 GeV in
Eq. (\ref{f_P_log}).}
\end{figure}
\section{Summary}
To conclude, 
we find that the zero mode contribution to the pseudoscalar meson
propagators extends to a fairly long distance in the time separation.
We obtain the axial renormalization constant,
$Z_A = 0.1589(4)$, 
being fairly independent of $m_0 a$.
The renormalized $f_{\pi}$ gives $a = 0.148\, {\rm fm}$
and the physical lattice size is about
4 times the Compton wavelength of the lowest mass pion.
We have observed
quenched chiral log in the $f_P$ data set, with
$\delta$  about 0.4. 

This work is partially supported by DOE Grants DE-FG05-84ER40154 and 
DE-FG02-95ER40907.


\end{document}